# Mini Band Gap Generation in Magnetic Beta-Borophene: Effects of Optical Phonon Interaction

Defne Akay[1*], Santanu. K. Maiti[2]

[1]*Department of Physics, Faculty of Science, Ankara University, Tandogan/Ankara-06 100, Turkey*
[*]dakay@science.ankara.edu.tr

[2]*Physics and Applied Mathematics Unit, Indian Statistical Institute, 203 Barrackpore Trunk Road, Kolkata-700 108, India*

**Abstract**

In this work we report for the first time, to the best of our concern, the tunable electronic properties of Beta-Borophene (BB) on the polar substrate ($ZrO_2$). We provide an analytical prescription for the calculation of ground state energy in presence electron-phonon (e-ph) interaction, within the framework of the Lee-Low-Pines theory. In the theoretical investigation of the polaron formation in BB, we describe its effective masses, polaronic band-gap, mobility, and Fermi velocities, which are different in each coordinate due to the out-of-plane buckling structure. We also analyze how the average effective mass and Fermi velocities of the charge carriers in the buckling structure are affected by the external magnetic field. It is shown that the polaronic energy becomes more effective in presence of a magnetic field, which we confirm through an analytical prescription. The characteristics of the evolution of average effective mass and average effective Fermi velocity in the presence of a magnetic field are critically discussed. We find that the average effective Fermi velocity is less sensitive, while the other one is greatly influenced by the magnetic field.

## I. Introduction

Recently, researchers are paying attentions in other low-dimensional materials rather than the most well-known example, graphene [1]. Because of the unique properties in electronics, optoelectronics, and magnetics, and also due to the structural similarity and proximity to graphene, the other related systems are expected to share some of their superior properties. The growing interest in this field is triggered by the current silicon-based technology together with the possibility of engineering novel quantum materials exhibiting exotic electronic properties. However, because of the different out-of-plane buckling degrees, borophene has different structural properties than graphene [2]. Some 2D materials can also be made from bulk materials without being layered, such as the compound of hafenene, 2D flat boron, and 2D GaN [3, 4]. Metallicity is the most well-known property of borophene when compared to other semimetals (e.g., silicene and graphene) or semiconductors (e.g., phosphorene). Unlike bulk boron allotropes, borophene exhibits metallicity, which is consistent with the expectations of a highly anisotropic 2D metal [5]. Also, the bonding between monoatomic lattices of boron atoms is more complex than that of carbon, e.g., B-B bonds with two or three centers can form.

Furthermore, the magnetic field effect on two-dimensional materials has piqued the interest of researchers due to its prospective uses in spintronics and data storage [6-10]. Compared to the analogous bulk materials, the eventual thinness drastically alters the physical properties. However, due to elaborate structural design and complex micro-nano fabrication, many 2D magnetic materials have been predicted theoretically, but only a few have been produced in laboratory [11–15]. Furthermore, the discovery of new 2D magnetic materials is still governed by trial and error methods after exfoliation from the parent bulk compounds. Graphene, h-BN, transition metal dichalcogenides, and others are non-magnetic or weakly magnetic among light element-based 2D materials, but by optimizing defects, substrate, or adsorbing hydrogen atoms, they can be utilized as suitable functional materials for spintronics applications [16–18]. Boron has only three valence electrons, resulting in the formation of multicenter B-B bonds and a rich polymorphism with high chemical complexity in borophenes [19]. Several borophenes have been successfully synthesized in the ultrahigh vacuum on Ag, Cu, or Al substrates [20–21]. Some of the outstanding properties of atomically thin borophenes are the appearance of superconductivity, massless Dirac fermions, improved transport, and to name a few [22–27].

Based on the results of the linear response theory, some authors assessed the quantum Hall and longitudinal conductivity of a Hall field. In monolayer graphene, the Hall field does not eliminate valley degeneracy [28]. Magnetoconductivity has been extensively studied in stannane [29], phosphorene [30–31], silicene [32,33], topological insulators [34–37], and molybdenum disulfide [38–39] besides graphene. In addition to the modulation-induced Weiss oscillations in 8-Pmmn borophene [39], several theoretical studies of magnetotransport properties in 2D Dirac materials with tilted Dirac cones have been conducted [40–42]. Few theoretical studies on optical features have recently been published, including anisotropic plasmons [43,44], Drude weight, and the implications of particle-hole symmetry-breaking in optical conductivity [45]. Electric and magnetic field effects have also been studied in the electron wave packet dynamics that predict the tilted anisotropic Dirac cones in borophene [46]. The borophene structure is made up of seven B36 units as described by the Ising model. The magnetic and thermodynamic properties of it have been studied using Monte Carlo simulation [47]. From the study of the electronic properties of borophene based on first-principles calculations [48], it has been confirmed that phonon frequency analysis supports the stability of the strained structures. Otherwise, the density-density response function, i.e., the polarization/Lindhard function [51], the oblique Klein tunneling [52], the optical conductivity [53], the electronic dynamics under the action of surface optical phonons [49], and the nonlinear Hall effect [54], have been reported using the low-energy effective Hamiltonian of one-atom-thick boron, i.e., 8-Pmmn borophene. Furthermore, first-principles calculations show that the charge transfer interaction between borophene and substrates occurs with a low degree of covalent bonding. This research paves the way for the fabrication of borophene-based devices in the future [55].

The *optical polaronic effects of borophene on its polar substrate in a magnetic field*, on the other hand, have yet to be investigated [56]. Focussing on the purpose of developing more 2D intrinsic magnetism that outperforms magnetic doped materials in terms of characteristics [57-58], here we investigate the low-energy effective Hamiltonian based on the Dirac system. Its evolutionary structure has been predicted and explored in the magnetic field, and the effects on the polaronic structure have been discussed methodically. The low boron mass density reactions in strong electron-phonon coupling on the magnetic monoatomic lattice of boron atoms, i.e., borophene, which causes phonon-mediated superconductivity. Furthermore, due to the powerful anisotropic structure of borophene, its electronic properties can effectively be examined for magnetic and polaronic applications.

Electronic structure can be determined by the shape of the Fermi surface which gets changed rapidly from one point to another. In this study, the calculations of the optical phonons of borophene reveal Kohn anomalies at high-symmetry points ($K$ and $K'$) in the Brillouin zone. The anomalous behavior of the phonon dispersion has a stronger influence on the electronic band structure of the symmetric Dirac point. Kohn displacements is evident which is in agreement with recent angle-resolved photoemission spectroscopy measurements [59]. The magnetic field effects on the carriers recombination process is confined by the calculation. In fact, carriers mobilities are not the dominant mechanism to explain the electronic structure. But, it has become the extremely significant parameter in the evaluation.

The key findings of our work are: (i) band gap generation and its manipulation in beta-borophene system by means of magnetic field and electron-phonon coupling, (ii) analytical prescription for the calculation of ground state energy even in the presence of e-ph interaction, which is one of the key aspects of our work as analytical result always provides better understanding than the numerical one, and (iii) the specific role of magnetic field on polaronic energy.

The rest of the paper is organized as follows. Section II explains the theoretical structure and low-dimensional effective Hamiltonian of $\beta$-borophene, as well as the electronic band structure of all models considered, which are based on the $k \cdot p$ type of Hamiltonian. Also, the formulation of magnetic fields and an optical polaronic model using the Fröhlich Hamiltonian are presented to diagonalize the synthesized model on the polar substrate $ZnO_2$. The Lee-Low-Pines (LLP) transformation is utilized in the presence of an external magnetic field. The corresponding results for the magnetic field effects on the optical-polaronic structure of beta-borophene are presented in Sec. III. Finally, we conclude our results in Sec. IV.

## II. Theoretical Model

The single-particle low-energy effective model Hamiltonian of tilted anisotropic Dirac cones is used as a starting point [49]. Under an external magnetic field in the z direction $\vec{B} = B\hat{z}$, the momentum operator is modified as $\vec{\pi} \to \vec{p} - e\Lambda$. The vector potential $\Lambda = \frac{B}{2}(-y, x, 0)$, induced by the magnetic field is taken in the symmetric gauge and $\vec{p}$ is the momentum operator. The Hamiltonian is written as a sum

$$\hat{H} = H_e + H_{ph} + H_{e-ph} \tag{1}$$

where

$$H_e = \begin{pmatrix} v_t \hat{\pi}_y + H & v_x \hat{\pi}_x - i v_y \hat{\pi}_y \\ v_x \hat{\pi}_x + i v_y \hat{\pi}_y & v_t \hat{\pi}_y \end{pmatrix}. \tag{2}$$

The direction-dependent velocity terms as reported in [49] are $v_x = 0.86\, v_F$, $v_y = 0.69\, v_F$, and $v_t = 0.32\, v_F$. Here $v_F$ is the Fermi velocity in the unit of $10^6$ m/s. Note that anisotropy term $v_x/v_y$ arises when not equals to 1, while $v_t$ ensures a tilt through non-concentric constant energy contours. It is known [49] that dispersion the relation in the absence of magnetic field $E_\pm(k) = \hbar k (v_t \sin\theta) \pm \sqrt{v_x^2 \cos^2\theta + v_y^2 \sin^2\theta}$. The conduction (valence) band in $\beta$-borophene is denoted by the upper (lower) sign. This fundamental energy description is used to investigate scenarios including an external magnetic field and polaronic interaction. This model assumes that a single layer of borophene is sandwiched between the substrate and the air, and that there is a strong interaction between the optical phonon on the surface of the substrate and the electronic (hole) in the borophene. The phonon part of the model can be stated as,

$$H_{ph} = \begin{pmatrix} \sum_{\rho}\sum_{\vec{q}} \hbar\omega_\rho b_{\vec{q}}^\dagger b_{\vec{q}} & 0 \\ 0 & \sum_{\rho}\sum_{\vec{q}} \hbar\omega_\rho b_{\vec{q}}^\dagger b_{\vec{q}} \end{pmatrix}. \qquad (3)$$

Here, $\omega_\rho$ is the surfaces optical (SO) phonon frequency with two branches ($\rho = 1,2$) and $b_{\vec{q}}^\dagger (b_{\vec{q}})$ is the creation (annihilation) operator for the SO phonons. Also, its interaction part Hamiltonian is as follows:

$$H_{e-ph} = \begin{pmatrix} \sum_{\rho}\sum_{\vec{q}}(M_{\vec{q}\rho}^* b_q^\dagger + M_{\vec{q}\rho} b_{\vec{q}}) & 0 \\ 0 & \sum_{\rho}\sum_{\vec{q}}\left(M_{\vec{q}\rho}^* b_{\vec{q}}^\dagger + M_{\vec{q}\rho} b_{\vec{q}}\right) \end{pmatrix}. \qquad (4)$$

Within the Lee-Low-Pines (LLP) model, the first and second unitary transformations are respectively,

$$U_1 = \exp\left[-i\vec{r}\cdot\sum_{\vec{q}} \vec{q}\, b_{\vec{q}}^\dagger b_{\vec{q}}\right].$$

The $U_1$ transformation removes the $\vec{r}$ electron coordinates in the Hamiltonian. Applying the transformation on $p_i$ and $b_{\vec{q}}$ as, $\tilde{p}_i = U_1^{-1} p U_1 = p_i - \sum_q b_{\vec{q}}^\dagger b_{\vec{q}}$ and $\tilde{b}_{\vec{q}} = U_1^{-1} b_{\vec{q}} U_1 = b_{\vec{q}} e^{i\vec{q}\cdot\vec{r}}$ respectively.

$$U_2 = exp\left[\sum_{\vec{q}}\left(f_{\vec{q}} b_{\vec{q}}^\dagger - f_{\vec{q}}^* b_{\vec{q}}\right)\right]$$

The second transformation adjusts the phonon coordinates, which allows us to account for the dressed electron states caused by the phonon field's coherent states. By the way, the optical phonon transformation is used to realize the form, $U_2^{-1} b_{\vec{q}} U_2 = b_{\vec{q}} + f_{\vec{q}}$. According to the transformation $f_{\vec{q}}$ is the variational function. Thus, by the successive two unitary transformations, phononic and electronic parts can be converted into,

$$\widehat{H}_p' = U_2^{-1} U_1^{-1}(H_{ph} + H_{e-ph}) U_1 U_2 \qquad (5)$$

$$\widehat{H}_e' = U_2^{-1} U_1^{-1} H_e U_2 U_1 . \qquad (6)$$

We modify the low-energy continuum two-band Hamiltonian to expand on these findings [12] for β-borophene in the presence of an external magnetic field and optical polaronic effect that describes an external magnetic field and an optical polaronic effect along the high-symmetry points in the Brillouin zone. The components of the transformed effective Hamiltonian can be expressed as

$$\widehat{H}' = \widehat{H}_e' + \widehat{H}_p' = \begin{pmatrix} H_{11}' & H_{12}' \\ H_{21}' & H_{22}' \end{pmatrix}.$$

It is also seen from the matrix elements, diagonal entries are equal but note that the off-diagonal entries are complex conjugate with respect to each other. This has made it easy for us to do our very complex transactions.

$$H'_{11} = H'_{22} = v_t \left[ \frac{\hbar\lambda}{\sqrt{2}}(b_y^\dagger + b_y) - \sum_{\vec{q}} \hbar\vec{q}\left(b_{\vec{q}}^\dagger + f_{\vec{q}}\right)\left(b_{\vec{q}} + f_{\vec{q}}^*\right) + \frac{eB}{2}\left(\frac{i}{\sqrt{2}\lambda}\right)(b_x - b_x^\dagger) \right]$$
$$+ \sum_{\vec{q}} \left[ \hbar\omega_\mu \left(b_{\vec{q}}^\dagger + f_{\vec{q}}\right)\left(b_{\vec{q}} + f_{\vec{q}}^*\right) + M_{\vec{q}}(b_{\vec{q}} + f_{\vec{q}}^*) + M_{\vec{q}}^*\left(b_{\vec{q}}^\dagger + f_{\vec{q}}\right) \right],$$

$$H'_{12} = v_x \left[ \frac{\hbar\lambda}{\sqrt{2}}(b_x^\dagger + b_x) - \sum_{\vec{q}} \hbar\vec{q}\left(b_{\vec{q}}^\dagger + f_{\vec{q}}\right)\left(b_{\vec{q}} + f_{\vec{q}}^*\right) - \frac{eB}{2}\left(\frac{i}{\sqrt{2}\lambda}\right)(b_y - b_y^\dagger) \right]$$
$$- iv_y \left[ \frac{\hbar\lambda}{\sqrt{2}}(b_y^\dagger + b_y) - \sum_{\vec{q}} \hbar\vec{q}\left(b_{\vec{q}}^\dagger + f_{\vec{q}}\right)\left(b_{\vec{q}} + f_{\vec{q}}^*\right) + \frac{eB}{2}\left(\frac{i}{\sqrt{2}\lambda}\right)(b_x - b_x^\dagger) \right],$$

$$H'_{21} = v_x \left[ \frac{\hbar\lambda}{\sqrt{2}}(b_x^\dagger + b_x) - \sum_{\vec{q}} \hbar\vec{q}\left(b_{\vec{q}}^\dagger + f_{\vec{q}}\right)\left(b_{\vec{q}} + f_{\vec{q}}^*\right) - \frac{eB}{2}\left(\frac{i}{\sqrt{2}\lambda}\right)(b_y - b_y^\dagger) \right]$$
$$+ iv_y \left[ \frac{\hbar\lambda}{\sqrt{2}}(b_y^\dagger + b_y) - \sum_{\vec{q}} \hbar\vec{q}\left(b_{\vec{q}}^\dagger + f_{\vec{q}}\right)\left(b_{\vec{q}} + f_{\vec{q}}^*\right) + \frac{eB}{2}\left(\frac{i}{\sqrt{2}\lambda}\right)(b_x - b_x^\dagger) \right].$$

$\hat{H}'$ is diagonalized by squaring it, and its expectation values may be found using $E_{B,n}^2 = \langle 0|\langle \psi_n|H_e'^2|\psi_n\rangle|0\rangle$. To obtain the eigenvalues, a long and complex procedure is needed to achieve the result in terms of diagonal matrix components which are given as

$$H_e'^2 = \begin{pmatrix} H_{11}'^2 + H_{12}'H_{21}' & H_{11}'H_{12}' + H_{12}'H_{22}' \\ H_{11}'H_{21}' + H_{21}'H_{22}' & H_{22}'^2 + H_{12}'H_{21}' \end{pmatrix}. \tag{7}$$

In which the relations $b_i|n\rangle = \sqrt{n}|n-1\rangle$ and $b_i^\dagger|n\rangle = \sqrt{n+1}|n+1\rangle$ correspond to coordinate dependent relations. Additionally, $b_q|0\rangle = 0$ depends on phonon wave vector.

In the presence of a magnetic field, the eigenfunction of the system may be expressed as

$$|\Psi_n\rangle|0\rangle = \frac{1}{\sqrt{2}}\begin{pmatrix} C_n'|n-1\rangle|0\rangle \\ C_n|n\rangle|0\rangle \end{pmatrix}. \tag{8}$$

together with

$$C_n' = 1 - \delta_{n,0}$$
$$C_n = \sqrt{(1 + \delta_{n,0})}. \tag{9}$$

Here, the parameters are carrier eigenfunction and $|0\rangle$ describe the ground state i.e., zero phonon states. $f_{\vec{q}} = -\frac{M_{\vec{q},\rho}}{\hbar\omega_f}$ and $f_{\vec{q}}^* = -\frac{M_{\vec{q},\rho}^*}{\hbar\omega_f}$ are the variational parameters which allows to minimize the total energy of the system. $f_q$ and $f_q^*$ are complex conjugate with respect to each other. The parameters depend on the interaction frequency which can be written as, $\omega_f = \alpha\omega_\rho - \frac{\bar{v}}{\bar{\omega}_-}[2\sum_{\vec{q}}(\eta\vec{q}.\vec{k} + \vec{q}^2)]$. Here, $\bar{\omega}_+/\bar{\omega}_- = \alpha$ and its frequencies dependence $\bar{\omega}_+ = \omega_\rho + v_t(2\eta\vec{k} + \vec{q})$, $\bar{\omega}_- = -\omega_\rho + v_t(2\eta\vec{k} + \vec{q})$ and as well as the velocity dependence $\bar{v} = v_x^2 + v_y^2 + v_t^2$ which all appear when seen the variational operators. The interaction amplitude of electrons with SO-phonons of the substrate also defines the variational parameters, and its spatial dependence is provided by

$$M_{q,\rho} = w_{SO,\rho} \frac{e^{-qz}}{\sqrt{q}} \tag{10}$$

where $w_{SO,\rho} = \sqrt{e^2 \chi \hbar \omega_{SO,\rho}/2\varepsilon_0}$ is the coupling parameter and $z$ is the distance of the electron from the surface of the substrate. In this study, all analysis have been done setting the distance $z = 2$ Å. SO-phonon modes for different substrates is known [60]. To evaluate the surfaces interactions, in this study a polar substrate $ZrO_2$ has been used because $ZrO_2$-induced polaronic contribution is greater than that of other known polar substrates. For the $ZrO_2$, SO phonon modes are known those are $\omega_{SO,1} = 25\ meV$ and $\omega_{SO,2} = 71\ meV$, respectively. $\chi$ is the combination of the known $\kappa_\infty = 4\ \varepsilon_0$, $\kappa_0 = 24\ \varepsilon_0$, low-and high-frequency dielectric constants of the substrates which may be written as $\chi = (\kappa_0 - \kappa_\infty)/((\kappa_\infty + 1)(\kappa_0 + 1))$. Now, the only preferred direction in the system is the direction of momentum vector, i.e., $\vec{k}$, so because of the symmetry rules $\sum_{\vec{q}} \vec{q}\ |f_{\vec{q}}|^2$ should be differ from $\vec{k}$ by a scalar, $\sum_{\vec{q}} \vec{q}\ |f_{\vec{q}}|^2 = \eta \vec{k}$. A solution can be easily verified for $\eta$ and the results can be given in terms of scalar value of $\eta$. Also, via $\eta$ we can allow the minimization of total energy and it is possible to calculate the SO phonon's eigenenergy and coupling terms. Finally, in terms of diagonalized Hamiltonian parameters, the total energy may be represented as,

$$\begin{aligned} E_{B,n}^2 = \frac{1}{2} [ & C_n'^2 \langle 0|\langle n-1|(H_{11}^2 + H_{12}H_{21})|n-1\rangle|0\rangle \\ & + C_n' C_n \langle 0|\langle n-1|(H_{11}H_{12} + H_{12}H_{21})|n\rangle|0\rangle \\ & + C_n' C_n \langle 0|\langle n|(H_{21}H_{11} + H_{22}H_{21})|n-1\rangle|0\rangle \\ & + C_n^2 \langle 0|\langle n|(H_{22}^2 + H_{12}H_{21})|n\rangle|0\rangle] \end{aligned} \tag{11}$$

In the compact form corresponding energy eigenvalues of the system can be written as

$$E_{B,0}^2 = \frac{1}{2} \left\{ \left(\sqrt{2}\right)^2 \langle 0|\langle n|(H_{22}^2 + H_{12}H_{21})|n\rangle|0\rangle \right\}. \tag{12}$$

The expectation value of the energy of the polaron and magnetic field can be obtained via

$$\begin{aligned} E_{B,0} = \Bigg\{ & \bar{v} \left[ \left(\frac{\hbar \lambda}{\sqrt{2}}\right)^2 - \left(\frac{eB}{2}\right)^2 \left(\frac{i}{\sqrt{2}\lambda}\right)^2 + (\hbar \eta \vec{k})^2 + (\hbar^2 \eta \vec{q}.\vec{k}) \right] \\ & + \sum_{\vec{q}\rho} [\hbar \omega_\rho - \hbar v_t(\vec{q} + 2\eta \vec{k})] \left( M_{\vec{q}} f_{\vec{q}}^* + M_{\vec{q}}^* f_{\vec{q}} \right) + \\ & + \sum_{\vec{q}\rho} \left[ (\hbar \omega_\rho)^2 - 2\hbar v_t(\hbar \omega_\rho)(\vec{q} + \eta \vec{k}) \right] |f_{\vec{q}}|^2 + \sum_{\vec{q}} (\hbar \omega_\rho)^2 |f_{\vec{q}}|^4 \Bigg\}^{1/2}. \end{aligned} \tag{13}$$

Polaronic and magnetic corrections on energy can be seen by

$$\Delta E = E_{B,0} \pm E_\pm(k). \tag{14}$$

There are some important remarks concerning this energy equation. It describes an energy contribution from the optical phonon vibration and an external magnetic field. The energy is a result of vibrational activity and is related to the dynamical band gap that induces a Kohn anomaly.

It should be noted that the energy caused by the optical absorpsion is proportional to the cutt-off wave vector $q_{cut}$. The value of the $q_{cut}$ is used as 0.45 Å, which was calculated from [49]. In fact, the energy that is affected by the dielectric constant of the substrates is exactly proportional to the energy of the surface phonons together with the intensity of the external magnetic field. In monolayer $\beta$-borophene structure, charge carriers have different effective masses due to the anisotropy of the system along the $x$ and $y$ directions. Furthermore, effective mass of the borophene, which is central to the consideration of the optical properties of solids is very useful in parameterising the dynamics of band electrons when they are subjected to external forces. In this subsequent stage, the effective average masses can be evaluated by effective average mass,

$$\langle m^*_{i,j} \rangle = \langle \frac{1}{\hbar^2} \left( \partial^2 E / \partial k_i \, \partial k_j \right)^{-1} \rangle.$$

It can be observed that this is similar to Newton's second law of motion, with the exception that the real particle mass is substituted with an effective mass $m^*$. The effective mass is generally expressed in units of electron rest mass in a vacuum, i.e., $m^*/m_0$.

Effective mass has an important factor on the mobility of a charge carriers. Effective mass differences help to determine intrinsic mobility differences across semiconductors and variations between electron and hole conductions within a particular semiconductor. Furthermore, the effective mass is a constant for a given material but combined effects of optical phonon and external magnetic field will contribute its behavior. The energy dispersion relations are very important in the determination of the transport properties for the carriers in solids. In this part, average effective velocity represents the Fermi velocity renormalization in the presence of polaronic effects and magnetic field and can be written as,

$$\langle v_{i,j} \rangle = \langle \frac{1}{\hbar} \partial E / \partial k_{i,j} \rangle.$$

The equation can be discussed from a more advanced point of view which considers also the case of degenerate bands in the physics of solids sequence. The carrier mobility describes how fast an electron or hole can move when driven by a field, and it is heavily influenced by scattering processes. The scattering caused by phonons defines the "intrinsic" mobility, which establishes the upper limit for free-standing and defect-free 2D semiconductors. Due to the presence of different scattering contributions, such as defects, it is quite difficult to measure the carrier mobility experimentally, but the intrinsic mobility is more often studied theoretically.

Based on the quantum statistics theory [61,62], at finite temperature, the mean number of the phonons is expressed as $\overline{N} = [exp(\hbar\omega_\rho/k_B T) - 1]^{-1}$. The mobility is inversely proportional to the mean number of phonons that is $\mu \approx 1/\overline{N}$. Here, $k_B$ and $T$ are the Boltzmann constant and the temperature of the monolayer polaronic $\beta$-borophene system, respectively.

As an extrinsic factor, scattering from substrate dielectrics is also an important issue. The phonons in the dielectric material can help mobility, especially if the semiconductor and the dielectric have a strong interaction or if the dielectric has polar vibrational modes (which can induce long-range Coulomb scattering). Experiments have shown that different dielectrics can result in varying measured mobilities. Relying on the calculations, we show that carrier transport in borophene at low carrier concentration occurs by small polaron hopping to evaluate the temperature dependence of mobility. Additionally, as a substrate, we used a zirconium dioxide ceramic structure, which is a polar material. Because the $ZrO_2$ induced polaronic contribution is higher than other substrates [49].

To examine the intrinsic mobility and Fermi velocity, the interaction between the material and selected substrate needs to be considered and it plays a crucial role on the physical properties. Borophene/$ZrO_2$ shows high electron-phonon interaction via polar coupling to optical phonons, resulting the production of tiny polarons. Some researchers are investigating to explore whether the mobility is related to the electronic structure properties such as band gap and effective mass. For example, Cheng and Liu [64] have presented a density functional perturbation method to evaluate this scenario. They have shown that the mobility does not correlate with the band gap or the effective mass. The uncorrelated relationship between mobility and effective mass calls into question the widely held belief that a low effective mass usually trades off high mobility. As a result, a material with a low effective mass (suggesting a high carrier concentration) but high mobility and thus a high current density can also obtained.

Based on the above theoretical framework and analytical prescription, we have determined our results. In a comprehensive way, we have analyzed the mobility of tiny polarons in a relevant temperature range. The effect of temperature on band transport has been critically investigated. Additionally, the polaronic band gap and the effects of increasing the magnetic field on the movement of carriers have been considered. Finally, the characteristics of the evolution of an average effective mass and an average effective Fermi velocity in the presence of a magnetic field have been carefully studied.

### III. Results and Discussion

To obtain greater insight into the electronic properties of the $\beta$-borophene, the polaron effective mass, renormalized average velocity and energy gap are significant quantities that can serve as a connection between experimental data and theoretical research in which all assumptions are established. In the present paper we have calculated the effective mass of the $\beta$-borophene interacting Fröhlich polarons, which are adapted from the Lee-Low-Pines theory. We have evaluated the average effective mass according to its change with respect to the magnetic field and the optical polaronic effect. It is obvious from these results that equation (14) together with equation (13) not only provide the results for the ground-state binding energy of an optical polaronic effect in anisotropic 2D monolayer borophene but also improve the understanding of their behavior in the wide range of magnetic strengths. To reveal this fact, in Fig. 1 we have presented a comparison between the magnetic field and $k$-dependence of corrected energy that have been obtained within the variational method. It can be clearly seen from the figure that (i) the application of a uniform magnetic field yields a further enhancement of the band gap energy, and (ii) as the strength of the magnetic field increases, the corrections on the band gap become quite large as compared to those obtained without magnetic field. Another important feature revealed from the figure is the fact that the introduction of the $k$-dependence of the energy spectrum has a minor effect on the magnetic field. Furthermore, the $k$-dependent effects get decreased more by enhancing the strength of magnetic field.

In Fig. 2(a), we have plotted the average effective mass to the band gap energy of the ground state in the monolayer structure on the $ZrO_2$ characteristic frequency and on its effective screening parameters. It is shown that the effective mass of the structure increases with increasing the magnetic field strength in the intermediate coupling region.

The interaction with optical phonons leads to the formation of a dressed electronic state at the energy level. The energy level increases by increasing the magnetic field. This can be attributed to a more number of virtual phonons accompanying the electron with increasing magnetic field. In this study, spatial dependencies of the velocity are ignored. The value of Fermi velocity in the structure is similar to that of the Fermi velocity in graphene [65]. From Fig. 2(b) it can be seen that the effective average velocity of the anisotropic structure feels the effective magnetic field. In this case, average velocity value gets lowered than that of the without magnetic field state. This can be attributed to a virtual magnetic barrier on the polaronic structure.

Finally, the variation of mobility is shown in Fig. 3. The lattice contribution is determined by the atomic density, whereas the electronic contribution is evaluated from the conductivity. The conductivity is related to the lattice vibration and the electronic contributions. Here it is important to note that the electrical conductivity depends on the mobility, and the mobility, on the other hand, is related to the effective mass and the mean free time between collisions of electrons that cause the scattering. So, it is expected that the mobility of the structure will increase with increasing the magnetic field. The effect of temperature is also crucial. With rising the temperature the magnetic contribution towards the polaronic mobility gets scaled up.

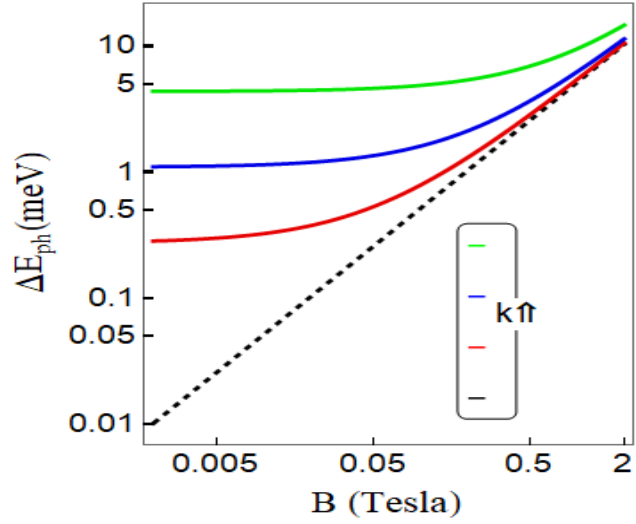

Fig. 1 (Colour online). Change of energy contribution with magnetic field for different values of $k$, where $k = 0$ (dashed black), $0.001$ (red), $0.002$ (blue), and $0.004$ (green).

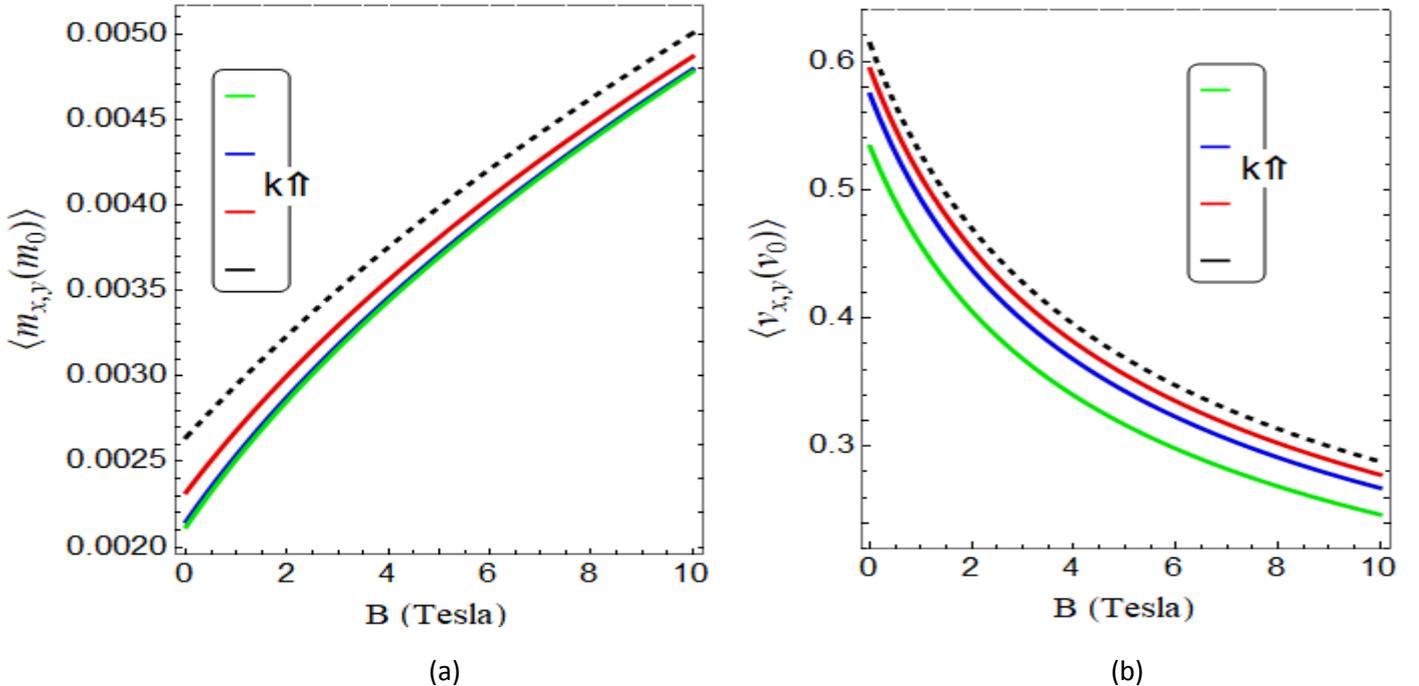

(a)                                    (b)

Fig. 2: (Colour online). (a) Variation of average effective mass with magnetic field B. (b) Dependence of average effective Fermi velocity with magnetic field B. In both (a) and (b), the results are evaluated for identical set of $k$ values where $k = 0$ (dashed black), $0.001$ (red), $0.002$ (blue), and $0.004$ (green).

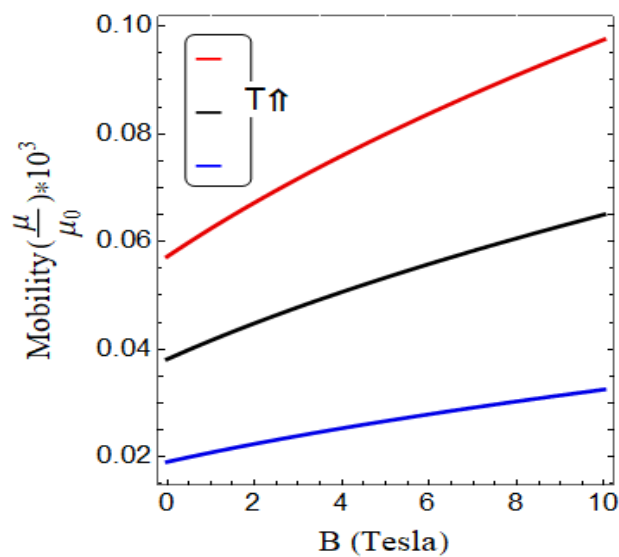

Fig. 3: (Colour online). Variation of mobility with magnetic field B at three typical temperatures, where the red, black and blue curves are associated with 300K, 200K and 100K, respectively.

## IV. Closing Remarks

We have investigated the tunability of the electronic characteristics caused by the external magnetic field and an optical phonon in $\beta$-borophene in presence of a polar substrate. We have provided an analytical approach for calculating the ground state energy of the electron-phonon system within the framework of the Lee-Low-Pines theory. We have discussed its renormalized effective masses and Fermi velocities, which differ in each coordinate due to the out-of-plane buckling structure, in the theoretical research of polaron production in $\beta$-borophene. Here, we have ignored the spatial dependencies of the effective velocity and mass. In contrast to the effective average velocity, the effective average mass is an increasing function of the magnetic field. For a given temperature range, the mobility gets enhanced with the rise of temperature. In particular, this shows that the structural model can be modified by external fields and intrinsic characteristics. Our present analysis may provide some key concepts for further research work considering the borophene system in designing 2D energy storage materials as well as supercapacitors.